%% file: main.tex
\documentclass[10pt, conference, letterpaper]{IEEEtran}
\IEEEoverridecommandlockouts %

\usepackage{cite}
\usepackage[hyphens]{url}

\usepackage{graphicx}
\usepackage{subfigure}
\usepackage{algorithm}
\usepackage[noend]{algorithmic}
\usepackage{amssymb,amsmath,graphicx}
\usepackage{bm}
\usepackage{enumerate}
\usepackage{xspace}
\usepackage{siunitx}
\usepackage[short]{optidef}

\usepackage{mathtools}
\usepackage{tabu}
\usepackage{epstopdf}
\usepackage[final,commentmarkup=margin]{changes}
\definechangesauthor[name={Tao Zhao}]{Tao}
\setdeletedmarkup{} %

\newcommand{\new}{}

\newtoggle{arxiv}
\toggletrue{arxiv}

\renewcommand*{\vec}[1]{\bm{#1}}

\newcommand{\goto}{\rightarrow}

\newtheorem{lemma}{Lemma}

\newtheorem{theorem}{Theorem}

\newtheorem{remark}{Remark}

\newcommand*{\cave}{CaVe\xspace}
\newcommand*{\cop}{CoP\xspace}
\newcommand*{\cavecop}{\cave-\cop}

\begin{document}
\title{Cache-Version Selection and Content Placement for\\ {Adaptive} Video Streaming in\\ {Wireless Edge Networks}}
\newcommand{\CoFirstAuthor}{\IEEEauthorrefmark{1}}
\author{\IEEEauthorblockN{Archana Sasikumar\CoFirstAuthor\IEEEauthorrefmark{2}, Tao Zhao\CoFirstAuthor\IEEEauthorrefmark{3}, I-Hong Hou\IEEEauthorrefmark{3}, and Srinivas Shakkottai\IEEEauthorrefmark{3}}%
\IEEEauthorblockA{\IEEEauthorrefmark{2}Juniper Networks}
\IEEEauthorblockA{\IEEEauthorrefmark{3}Dept. of ECE, Texas A\&M University, College Station, TX 77843\\
Email: asasi@juniper.net, \{alick,ihou,sshakkot\}@tamu.edu}%
\thanks{\CoFirstAuthor These authors contributed equally to this work.}%
\thanks{A.~Sasikumar was with Texas A\&M University when conducting this work.}%
\thanks{\added{This research was supported in part by grants NSF CNS 1149458,
  AST 1443891, NSF-Intel CNS 1719384, ARO W911NF-18-1-0331, and ONR
N00014-18-1-2048.}}
}
\maketitle

\begin{abstract}
Wireless edge networks are promising to provide better video streaming
services to mobile users by provisioning computing and storage resources at
the edge of wireless network. However, due to the diversity of user interests,
user devices, video versions or resolutions, cache sizes, network conditions,
etc., it is challenging to decide where to place the video contents, and which
cache and video version a mobile user device should select. In this paper, we
study the joint optimization of cache-version selection and content placement
for adaptive video streaming in wireless edge networks. We propose practical
distributed algorithms that operate at each user device and each network cache
to maximize the overall network utility. In addition to proving \replaced{the
optimality of our algorithms}{that our algorithms indeed achieve the optimal
  performance}, we implement our algorithms as well as several baseline algorithms on ndnSIM, an ns-3 based Named Data Networking simulator. Simulation evaluations demonstrate that our algorithms significantly outperform conventional heuristic solutions.
\end{abstract}
\maketitle

\newcommand{\etal}{\textit{et al.}\xspace}
\input{intro.tex}

\input{system.tex}

\input{cv.tex}

\input{cp.tex}

\input{implementation.tex}

\input{simulation.tex}

\input{related.tex}

\input{conclusion.tex}

\end{document}

%% file: intro.tex
\section{Introduction}
\label{sec:intro}

Video streaming has become the dominant application for modern Internet traffic. In order to provide better \added{quality} of service (QoS) and quality of experience (QoE) to mobile users, content delivery networks (CDNs) have been deployed to store popular videos at cache servers close to the users. This aligns with the trend of wireless edge networks, where computing and storage resources are provisioned at the edge of the wireless network~\cite{mach2017mobile}. Meanwhile, as users are accessing videos from a variety of devices, ranging from smartphones to 4K televisions (TVs), adaptive video streaming, which encodes the same video content into multiple versions with different resolutions, has been widely used to deliver arguably the best video version to each user based on device types and network conditions.  %

In this paper, we study the interplay between three important components for
adaptive video streaming in wireless edge networks: \emph{cache selection},
where each user device determines which cache server to retrieve videos from,
\emph{version selection}, which determines the version that each user watches,
and \emph{content placement}, which entails the caching strategy of each cache
server. We formulate CaVe-CoP, a Cache-Version selection and Content Placement
problem that jointly optimizes these three components by taking into account
the preferred video versions of users, the communication capacities of network
links, and the storage capacities of cache servers.
Our goal is to develop a new network algorithm for CaVe-CoP that is not only provably optimal, but also practical and implementable.

Our proposed solution is based on the observation that there is a practical
timescale separation between cache-version selection (CaVe) and content
placement (CoP), as the former can be updated much more frequently. Hence, we
first solve the CaVe problem by fixing the solution to the CoP problem\added{,
and prove the optimality of our CaVe algorithms}. We then solve the CoP
problem by considering its influence to solution to the CaVe problem\added{,
  and prove our CoP algorithms are optimal when fractional solutions are
allowed}.\deleted{ For both problems, we propose practical distributed
algorithms and prove that they converge to the optimal solutions.}

\replaced{While our algorithms can be practically implemented under the current Internet
architecture with TCP/IP, we demonstrate that our algorithms can also be implemented
in a distributed fashion on Named Data Networking
(NDN)~\cite{zhang2014ndn}, a future Internet architecture designed with
video streaming applications in mind.}{%
We further implement our algorithms on Named Data Networking
(NDN)~\cite{zhang2014ndn}, a future Internet architecture where network
routers have built-in caches, and evaluate our algorithms on
ndnSIM~\cite{mastorakis2017ndnsim}, an ns-3 based NDN simulator. We
demonstrate that our algorithms can be implemented in a distributed fashion.}
Since NDN forwards packets by content names instead of location IDs such as IP
addresses, we present a distributed forwarding strategy that ensures user
devices always obtain their selected video versions from their selected cache
server. Moreover, we show that the overhead of our algorithms is negligible by
exploiting local information and built-in caching. \added{We evaluate our algorithms on
ndnSIM~\cite{mastorakis2017ndnsim}, an ns-3 based NDN simulator. }Simulation results depict that our algorithms significantly outperform baseline policies that employ conventional heuristic solutions and subsets of our algorithms.

The rest of the paper is organized as follows. Section~\ref{section:model} introduces our system model and the formulation of CaVe-CoP. Solutions to the two problems CaVe and CoP are introduced in Section~\ref{section:cache_version} and ~\ref{section:cp}, respectively. In Section~\ref{section:implementation}, we discuss the implementation of our algorithms in NDN. Section~\ref{sec:sim} demonstrates the simulation results. Section~\ref{sec:related} reviews some related literature. Finally, Section~\ref{section:conclusion} concludes the paper.

%% file: system.tex
\section{System Model}
\label{section:model}

\begin{figure}[!t]
  \centering
  \includegraphics[width=0.4\textwidth]{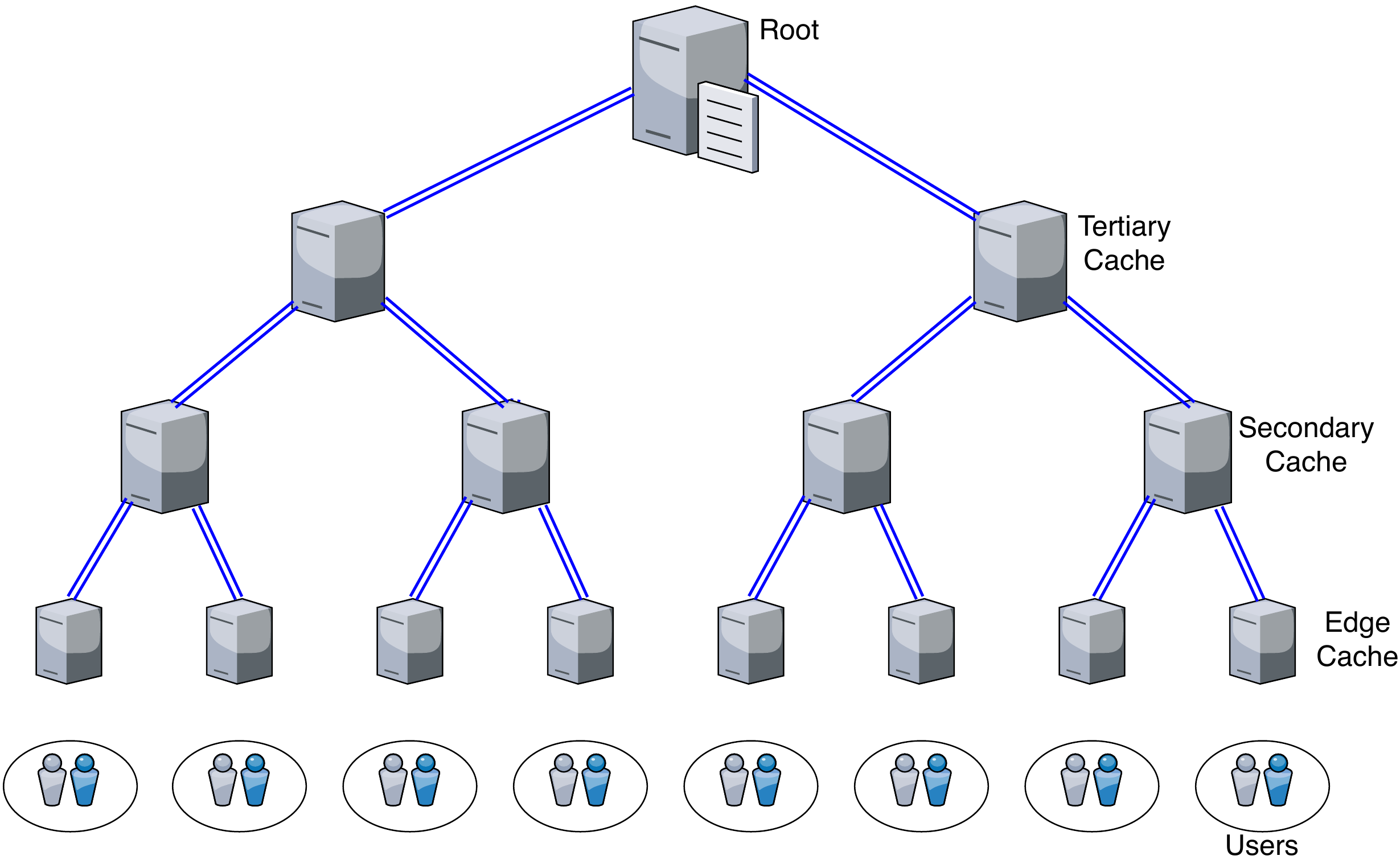}
   \caption{A wireless edge network with a root node holding all videos and three layers of caches. Each edge cache serves a group of users with different user devices.}
   \label{fig:topo}
\end{figure}

We consider a wireless edge network where a group of network caches jointly host a set of videos and serve a set of video streaming users.\footnote{The terms ``user'' and ``user device'' are used interchangeably.} Fig.~\ref{fig:topo} illustrates an example of such a network, which is consistent with the YouTube video delivery system~\cite{ramadan2017big}.
We use $\mathbb{C}$ to denote the set of network caches, $\mathbb{S}$ to denote the set of users, and $\mathbb{L}$ to denote the set of communication links that connect the network caches, routers, and users.  
We assume there is a route\footnote{Our model and algorithms can be  generalized to the multi-route scenario.} between each user $s$ and each network cache $c$, and define $H^l_{s,c}$ as the indicator function that link $l$ is on the route between $s$ and $c$.

We consider multi-version video streaming where each video is encoded into multiple versions for different resolutions of the same video content. We use $\mathbb{V}$ to denote the set of all versions of all videos. For each video version $v\in \mathbb{V}$, we use $X_v$ to denote the average bit rate of $v$ and $Y_v$ to denote the file size of $v$, i.e. the product of $X_v$ and the duration of the video. For the ease of theoretical analysis, we assume that there exists a \emph{null version} $v_0$ with $X_{v_0}=0$ and $Y_{v_0}=0$. If a user decides not to watch any video, then we say that the user watches the null version $v_0$. With the introduction of the null version, we can assume that each user always watches a video version.

Each network cache $c\in\mathbb{C}$ has a storage of size $B_c$ to store some video versions. Specifically, let $p_{c,v}$ be the indicator function that $v$ is present in the storage of $c$, then we have $\sum_v Y_vp_{c,v}\leq B_c$, for all $c$. Each network cache $c$ determines which video versions to store, and thereby determines the values of $p_{c,v}$, subject to its storage constraint. We assume that there exists at least a network cache $c$  with infinite storage $B_c=\infty$ and stores all video versions, which we call the root node. Such an assumption is to ensure that at least one copy of each video version exists in the network. We refer to the problem of determining $p_{c,v}$ as the \emph{content placement (CoP) problem}.

At the user end, each user $s$ is interested in watching a video. Let $\mathbb{I}_s$ be the set of video versions that correspond to the interested video of user $s$. Each user device $s$ needs to determine which video version to watch, as well as which network cache to obtain the video version from. Let $z_{s,c,v}$ be the indicator function that user $s$ decides to watch video version $v$, and to obtain it from network cache $c$. We refer to the problem of determining $z_{s,c,v}$ as the \emph{cache-version selection (CaVe) problem}. Since user $s$ needs to obtain exactly one video version, we require that $\sum_{c, v\in\mathbb{I}_s}z_{s,c,v}=1$, for all $s$. Moreover, user $s$ can only obtain video version $v$ from network cache $c$ if $c$ indeed stores $v$, that is, $p_{c,v}=1$. Hence, we also need $z_{s,c,v}\leq p_{c,v}$, for all $s,c,v$. 

Recall that the bit rate of video version $v$ is $X_v$ and $H^l_{s,c}=1$ if link $l$ is on the route between $s$ and $c$. When user $s$ obtains $v$ from $c$, it incurs an amount of $X_v$ traffic on each link along the route between $s$ and $c$. The total amount of traffic on link $l$ can then be expressed as $\sum_{s,c,v}X_vH^l_{s,c}z_{s,c,v}$. We consider that each link $l$ has a finite capacity of $R_l$, and hence we require that $\sum_{s,c,v}X_vH^l_{s,c}z_{s,c,v}\leq R_l$, for all $l\in\mathbb{L}$.

Finally, each user obtains some utility based on its perceived video quality. In particular, we consider that each user $s$ obtains a utility of $U_s(X_v)$ when watching a video version with bit rate $X_v$. We assume that $U_s(\cdot)$ is a non-decreasing and concave function. Different users may have different utility functions since they may be watching videos on different types of devices. For example, users watching videos on smartphones typically enjoy lower utility than those watching videos on TVs.

We aim to maximize the total utility of all users in the network by choosing the optimal $\vec{p}:=[p_{c,v}]$ and $\vec{z}:=[z_{s,c,v}]$, subject to all aforementioned constraints. Formally, we have the following CaVe-CoP optimization problem.\footnote{In practice the CaVe-CoP problem will be solved repeatedly over time with different parameters to cope with network changes.}

\textbf{CaVe-CoP}
\begin{maxi!}[2]
  {}{\sum_{s,c,v\in \mathbb{I}_s}U_s(X_v)z_{s,c,v}\label{eq:NUM1}}{\label{eq:cavecop}}{}
  \addConstraint{\sum_vY_vp_{c,v}}{\leq B_c,\label{eq:NUM2}}{\forall c\in\mathbb{C}}
  \addConstraint{\sum_{c,v\in\mathbb{I}_s}z_{s,c,v}}{=1,\label{eq:NUM3}}{\forall s\in\mathbb{S}}
  \addConstraint{z_{s,c,v}}{\leq p_{c,v},\label{eq:NUM4}}{\forall s\in\mathbb{S}, c\in\mathbb{C}, v\in\mathbb{V}}
  \addConstraint{\sum_{s,c,v}X_vH^l_{s,c}z_{s,c,v}}{\leq R_l,\label{eq:NUM5}}{\forall l\in\mathbb{L}}
  \addConstraint{p_{c,v}\in\{0,1\}, z_{s,c,v}\in\{0,1\},}{\quad\label{eq:NUM6}}{\forall s\in\mathbb{S}, c\in\mathbb{C}, v\in\mathbb{V}.}
\end{maxi!}

While the utility maximization problem studied in this paper may look similar to many existing studies on network utility maximization (NUM), we note that there are two major challenges that distinguish our problem from other NUM problems: First, most existing studies on NUM problems assume that the source and destination of each flow is fixed and given. In contrast, multiple network caches may store the same video version depending on the solution to the content placement problem. Hence, not only does a user have multiple choices of network caches to obtain the video version from, but the problem of selecting cache is fundamentally intertwined with the problem of content placement. Second, although the problem of version selection may seem to be a special case of the rate control problem, we note that the problem of version selection is fundamentally intertwined with the problem of selecting cache since each cache may only store a subset of versions for a given video. The possibility of placing different versions of the same video at different caches also distinguishes this work from some recent studies on throughput-optimal algorithms with caches.
Araldo~\etal~\cite{araldo2016representation} studied a similar problem to ours. However, they only derived heuristics without meaningful performance guarantees.

The decision variables in CaVe-CoP are $\vec{p}$ and $\vec{z}$. We note that there is a practical timescale separation between the update for $\vec{p}$ and that for $\vec{z}$. When a user device changes its values for $\vec{z}$ due to e.g. network congestion, it simply requests new packets from a different network cache and/or with a different video version. Hence, $\vec{z}$ can be updated rather frequently, for example, once every 100 milliseconds. On the other hand, when a network cache changes its values for $p_{c,v}$, it needs to obtain all video versions with $p_{c,v}=1$. Hence, $\vec{p}$ can only be updated infrequently.

Our proposed solution for CaVe-CoP is based on the observation of the timescale separation between the update for $\vec{p}$ and that for $\vec{z}$. In Section \ref{section:cache_version}, we will first consider the CaVe problem by finding the optimal $\vec{z}$ for given $\vec{p}$. Next, in Section \ref{section:cp}, we will consider the CoP problem. In order to find the optimal $\vec{p}$, we will introduce pseudo-variables $\vec{z}':=[z'_{s,c,v}]$ and $\vec{p}':=[p'_{c,v}]$ that are updated at the same frequency as $\vec{p}$ to address the issue with timescale separation.

Finally, we note that \cavecop is an integer programming problem since $p_{c,v}$ and $z_{s,c,v}$ are integers. To obtain tractable results, we will relax (\ref{eq:NUM6}) and allow $p_{c,v}$ and $z_{s,c,v}$ to be any real number between 0 and 1. As we will demonstrate in Section \ref{section:cache_version}, our solution to the CaVe problem will always yield integer values for $z_{s,c,v}$. We will also discuss how to obtain integer solutions for  $p_{c,v}$ in Section \ref{section:cp}.

%% file: cv.tex
\section{The Cache-Version Selection Problem (CaVe)}
\label{section:cache_version}

In this section, we study the CaVe problem. We consider that the contents that
each network cache store are given and fixed, and aims to determine both the
video version to watch and the network cache to obtain contents from for each
user. In terms of the optimization problem (\ref{eq:NUM1})--(\ref{eq:NUM6}),
we focus on finding the optimal $\vec{z}:=[z_{s,c,v}]$ to maximize total
utility in the \added{network} when $\vec{p}:=[p_{c,v}]$ is given and fixed.

\subsection{Overview of the Solution}

We begin by rewriting the optimization problem (\ref{eq:NUM1})--(\ref{eq:NUM6}) for the CaVe problem. Since $\vec{p}$ is given and fixed, constraint (\ref{eq:NUM2}) no longer applies. Further, we relax the constraint (\ref{eq:NUM6}) by allowing $z_{s,c,v}$ to be any real number between 0 and 1. The resulting optimization problem, which we call CaVe-Primal, can then be described as follows:

\textbf{CaVe-Primal}
\begin{maxi!}
{}{\sum_{s,c,v\in \mathbb{I}_s}U_s(X_v)z_{s,c,v}\label{eq:NUM7}}{\label{eq:cave-primal}}{} 	
\addConstraint{\sum_{c,v\in\mathbb{I}_s}z_{s,c,v}}{=1,\label{eq:NUM8}}{\forall s\in\mathbb{S}}	
\addConstraint{z_{s,c,v}}{\leq p_{c,v},\ \label{eq:NUM19}}{\forall s\in\mathbb{S}, c\in\mathbb{C}, v\in\mathbb{V}}
\addConstraint{\sum_{s,c,v}X_vH^l_{s,c}z_{s,c,v}}{\leq R_l,\label{eq:NUM9}}{\forall l\in\mathbb{L}}
\addConstraint{0 \leq z_{s,c,v}}{ \leq 1,\label{eq:NUM10}}{\forall s\in\mathbb{S}, c\in\mathbb{C}, v\in\mathbb{V}.}
\end{maxi!}

We will consider a dual problem to CaVe-Primal. We associate a Lagrange multiplier, $\lambda_{l}$, for each link capacity constraint (\ref{eq:NUM9}), for all $l \in \mathbb{L}$. Let $\vec{\lambda}:=[\lambda_l]$ be the vector of Lagrange multipliers. The Lagrangian is obtained as follows:
\begin{align}	
&L(\vec{z}, \vec{\lambda}) \nonumber\\
&:= \sum_{s,c,v\in \mathbb{I}_s} U_s(X_v)z_{s,c,v} - \sum_l\lambda_{l}\left(\sum_{s,c,v}z_{s,c,v}H^{l}_{s,c}X_v - R_{l}\right) \label{eq:CV-L}
\end{align}

The dual objective, $D(\vec{\lambda})$, is defined as the maximum value of $L(\vec{z}, \vec{\lambda})$ over $\vec{z}$ subject to the constraints (\ref{eq:NUM8}), (\ref{eq:NUM19}), and (\ref{eq:NUM10}).
We call the underlying optimization problem CaVe-Lagrangian. It can be written as follows:

\textbf{CaVe-Lagrangian}
\begin{maxi!}
{}{L(\vec{z}, \vec{\lambda}) \label{eq:NUM11}}{}{}
\addConstraint{\sum_{c,v\in\mathbb{I}_s}z_{s,c,v}}{= 1,\label{eq:NUM12}}{\forall s \in\mathbb{S}}
\addConstraint{z_{s,c,v}}{\leq p_{c,v},\quad\label{eq:NUM12a}}{\forall s\in\mathbb{S}, c\in\mathbb{C}, v\in\mathbb{V}}
\addConstraint{0 \leq z_{s,c,v}}{ \leq 1,\label{eq:NUM13}}{\forall s\in\mathbb{S}, c\in\mathbb{C}, v\in\mathbb{V}.}
\end{maxi!}

\begin{remark}
In defining the CaVe-Lagrangian problem, we only relax the link capacity constraint (\ref{eq:NUM9}), and keep other constraints (\ref{eq:NUM8}), (\ref{eq:NUM19}) and (\ref{eq:NUM10}) intact. This is because the link capacity constraint (\ref{eq:NUM9}) can be temporarily violated as packets that cannot be served immediately can be queued in the buffer. On the other hand, constraints (\ref{eq:NUM8}) and  (\ref{eq:NUM19}) need to be satisfied at all time in practical systems.
\end{remark}

 The dual problem is to minimize $D(\vec{\lambda})$ while ensuring that all Lagrange multipliers $\lambda_l$ are non-negative. We call this the CaVe-Dual and mathematically write it as:
 
\textbf{CaVe-Dual}
\begin{mini!}
{}{D(\vec{\lambda}) \label{eq:NUM14}}{}{}
\addConstraint{\lambda_l}{\geq 0,\label{eq:NUM15}}{\quad\forall \lambda_l \in \mathbb{L}.}
\end{mini!}

\begin{theorem}[Strong Duality] \label{theorem:CV_strong_duality}
CaVe-Primal and CaVe-Dual have the same optimal value.
\end{theorem}
\iftoggle{arxiv}{
\begin{IEEEproof}
\new
The objective function of \cave-Primal is a linear function of
  $\vec{z}$, and hence is concave. The set of
$\vec{z}$ that satisfies the three unrelaxed constraints,
namely, (\ref{eq:NUM8}), (\ref{eq:NUM19}), and (\ref{eq:NUM10}), is nonempty and convex.

Furthermore, the
relaxed constraint (\ref{eq:NUM9}) is
linear and thus convex. To get strict inequalities in
(\ref{eq:NUM9}), we can
set $z_{s,c,v}$ to be $1$ if $c$ is the root node and $v$ is
the null version, and $0$ otherwise.
It is straightforward to see that
\eqref{eq:NUM8},
\eqref{eq:NUM19}, and
\eqref{eq:NUM10} are satisfied, while
\eqref{eq:NUM9} is satisfied with strict inequalities.

Hence, this theorem holds
following Theorem 6.2.4 (Strong Duality Theorem) in \cite{bazaraa2013nonlinear}.
\end{IEEEproof}
}{
\begin{IEEEproof}
  \added{See \cite{sasikumar2019cache}.}
\end{IEEEproof}
}

Based on Theorem~\ref{theorem:CV_strong_duality}, we can solve the CaVe-Primal problem by solving CaVe-Dual. Solving CaVe-Dual involves two steps: First, for a given vector $\vec{\lambda}$, we need to find $D(\vec{\lambda})$ by solving CaVe-Lagrangian. Second, we need to find the optimal $\vec{\lambda}$ to solve CaVe-Dual. We introduce our solutions to these two steps below.

\subsection{The Solution to CaVe-Lagrangian}

We rewrite (\ref{eq:CV-L}) as:
\begin{align}
&L(\vec{z}, \vec{\lambda}) \nonumber\\
&= \sum_{s,c,v\in \mathbb{I}_s} U_s(X_v)z_{s,c,v} - \sum_l\lambda_{l}\left(\sum_{s,c,v}z_{s,c,v}H^{l}_{s,c}X_v - R_{l}\right) \nonumber\\
&=\sum_s \sum_{c,v\in \mathbb{I}_s}z_{s,c,v}\big(U_s(X_v)-X_v\sum_{l:H^{l}_{s,c}=1}\lambda_l\big)+\sum_l\lambda_lR_l
\end{align}

We note that the above expression provides a natural user-by-user
decomposition. Specifically, by defining \added{$\vec{z}_s$} as the vector
containing all $[z_{s,c,v}]$ for a given $s$, and \added{defining}
\begin{equation}
L_s(\vec{z_s}, \vec{\lambda}):= \sum_{c,v\in \mathbb{I}_s}z_{s,c,v}\big(U_s(X_v)-X_v\sum_{l:H^{l}_{s,c}=1}\lambda_l\big),
\end{equation}
\deleted{then} we have
\begin{equation}
L(\vec{z}, \vec{\lambda})=\sum_s L_s(\vec{z_s}, \vec{\lambda})+\sum_l\lambda_lR_l.
\end{equation}

As $\vec{\lambda}$ is given in CaVe-Lagrangian, the last term $\sum_l\lambda_lR_l$ is a constant. Hence, $L(\vec{z}, \vec{\lambda})$ is maximized if one can maximize $L_s(\vec{z_s}, \vec{\lambda})$ for each user $s$. Moreover, recall that $p_{c,v}$ is the indicator function that network cache $c$ stores video version $v$. Therefore, the constraint (\ref{eq:NUM12a}) is equivalent to saying that $z_{s,c,v}$ needs to be 0 if $p_{c,v}=0$. We can now define CaVe-User$_s$ as follows:

\textbf{CaVe-User$_s$}
\begin{maxi!}
{}{\sum_{c,v: v\in \mathbb{I}_s, p_{c,v}=1}z_{s,c,v}\big(U_s(X_v)-X_v\sum_{l:H^{l}_{s,c}=1}\lambda_l\big) \label{eq:CV-User1}}{}{}
\addConstraint{\sum_{c,v: v\in \mathbb{I}_s, p_{c,v}=1}z_{s,c,v}}{ = 1\label{eq:CV-User22}}
\addConstraint{0 \leq z_{s,c,v}}{ \leq 1,\label{eq:CV-User3}}{\quad\forall c\in\mathbb{C}, v\in\mathbb{V}.}
\end{maxi!}

It is clear that the optimal vector $\vec{z}$ that solves CaVe-User$_s$, for
all $s$, is also the optimal vector that solves CaVe-Lagrangian. 
\replaced{To solve CaVe-User$_s$, note}{Note} that
the only decision variable in CaVe-User$_s$ is the vector \added{$\vec{z}_s$},
\replaced{while}{and all other variables, including} $U_s(X_v), X_v$, and $\lambda_l$ are
\added{all} constant\added{s}\deleted{ given \added{$\vec{z}_s$}}. Hence, the
following algorithm solves CaVe-User$_s$: First, find $(c^*,v^*)$ that has the
maximum value of $U_s(X_v)-X_v\sum_{l:H^{l}_{s,c}=1}\lambda_l$ among all
$(c,v)$ with $v\in \mathbb{I}_s$ and $p_{c,v}=1$. Ties can be broken
arbitrarily. Second, set $z_{s,c^*,v^*}=1$, and $z_{s,c,v}=0$ for all other
$(c,v)$. Alg.~\ref{AlgCVS-User} summarizes the algorithm. We note that, even
though we have relaxed the constraint and allowed $z_{s,c,v}$ to be any real
number between 0 and 1, \replaced{the optimal solution produced by
Alg.~\ref{AlgCVS-User} is always an integer one.}{Alg.~\ref{AlgCVS-User} shows
that the optimal solution to CaVe-Lagrangian is always an integer solution.} Besides, note that $c^*$ and $v^*$ are updated iteratively as $\vec{\lambda}$ is updated. It means the cache-version selection of each user is dynamic and adaptive to the network congestion.

\begin{algorithm}[h]
\caption{CaVe-User$_s$ Algorithm}
\label{AlgCVS-User}
\begin{algorithmic}
\STATE Obtain $\vec{p}$ and $\vec{\lambda}$
  \STATE $z_{s,c,v}\leftarrow 0, \forall c,v$
  \STATE $(c^*,v^*) \leftarrow \arg\!\max_{c,v\in\mathbb{I}_s: p_{c,v} = 1} U_s(X_v)-X_v\sum_{l:H^{l}_{s,c}=1}\lambda_l$
  \STATE $z_{s,c^*,v^*} \leftarrow 1$
\end{algorithmic}
\end{algorithm}

\subsection{The Solution to CaVe-Dual}

Our solution to CaVe-Dual is shown in Alg. \ref{AlgCVS-Link}, where each link $l$ updates its own $\lambda_l$. We have the following lemma and theorem.

\begin{algorithm}[h]
\caption{CaVe-Link$_l$ Algorithm}
\label{AlgCVS-Link}
\begin{algorithmic}
\STATE $t\leftarrow 0$, $\lambda_l \leftarrow 0$
\WHILE{true}  
\STATE Obtain $\vec{z}$ from Alg. \ref{AlgCVS-User}
\STATE $\lambda_l \leftarrow \left[\lambda_l + h_t(\sum_{s,c,v}X_vH^l_{s,c}z_{s,c,v} - R_l) \right]^+$ 
  \STATE $t\leftarrow t+1$
\ENDWHILE
\end{algorithmic}

\end{algorithm}

\begin{lemma}
  \label{lem:cave-subgradient}
Given $\vec{\lambda}$, let $\vec{z}^*$ be the vector that solves
CaVe-User$_s$. Then \added{$\vec{g}:=[g_l]:=[R_l - \sum_{s,c,v}X_vH^l_{s,c}
z^*_{s,c,v}]$} is a subgradient of $D(\vec{\lambda})$.
\end{lemma}
\iftoggle{arxiv}{
\begin{IEEEproof}
  \new
  $\vec{z}^*$ solves CaVe-User$_s$ and thus solves CaVe-Lagrangian.
  By definition,
  $D(\vec{\lambda}) = L(\vec{z}^*,\vec{\lambda})$ for the given $\vec{\lambda}$.
  Therefore, for any $\tilde{\vec{\lambda}}:=[\tilde{\lambda}_l]$ where $\tilde{\lambda}_l\ge 0$,
  \begin{align*}
    D(\tilde{\vec{\lambda}}) - D(\vec{\lambda}) &\ge L(\vec{z}^*,\tilde{\vec{\lambda}}) -
    L(\vec{z}^*,\vec{\lambda})\\
    &= - \sum_l (\tilde{\lambda}_l-\lambda_l)
    \left(\sum_{s,c,v} z^*_{s,c,v} H^l_{s,c} X_v - R_l\right)\\
    &= (\tilde{\vec{\lambda}}-\vec{\lambda})^T \vec{g}.
  \end{align*}
  Hence, $\vec{g}$ is a subgradient of $D(\vec{\lambda})$.
\end{IEEEproof}
}{
\begin{IEEEproof}
  \added{See \cite{sasikumar2019cache}.}
\end{IEEEproof}
}
\begin{theorem}
Let $\{h_t\}$ be a sequence of non-negative numbers with $\sum_{t=0}^\infty h_t=\infty$ and $\lim_{t\rightarrow\infty}h_t=0$, then Alg. \ref{AlgCVS-Link} solves CaVe-Dual.
\label{thm:cave-dual}
\end{theorem}
\iftoggle{arxiv}{
\begin{IEEEproof}
  \added{%
    Note that the objective function of CaVe-Dual is convex in
    $\vec{\lambda}$, and the feasible region is a nonempty, convex, closed
    subset of $\mathbb{R}^{|\mathbb{L}|}$.
    With Lemma~\ref{lem:cave-subgradient} and the step size sequence specified in the
    theorem, Alg.~\ref{AlgCVS-Link} solves CaVe-Dual following Theorem~8.9.2
    in~\cite{bazaraa2013nonlinear}.
  }
\end{IEEEproof}
}{%
\begin{IEEEproof}
  \added{See \cite{sasikumar2019cache}.}
\end{IEEEproof}
}

\subsection{\added{The Solution to CaVe-Primal}}

\added{We have the following theorem regarding the optimality of $\vec{z}$
  obtained by running
Alg.~\ref{AlgCVS-User} and Alg.~\ref{AlgCVS-Link} iteratively:}

\begin{theorem}
  \added{Let $\vec{z}^*$ be the vector that solves CaVe-Primal,
    $\vec{\lambda}^k$ be the vector produced by Alg.~\ref{AlgCVS-Link} after
    $k$ iterations, and
    $\vec{z}^k$ be the vector produced by Alg.~\ref{AlgCVS-User} when
    $\vec{\lambda}=\vec{\lambda}^k$.
    Let $\bar{\vec{z}}^t$ be the weighted average of $\vec{z}^k$ after the
    first $t$ iterations, i.e.
    $\bar{\vec{z}}^t := \lim_{T\goto\infty} \frac{\sum_{k=t+1}^{t+T} h_k
    \vec{z}^k}{\sum_{k=t+1}^{t+T} h_k}$.
    Then, for any $\varepsilon > 0$, there exists an integer $K$ such that for
    every $t>K$,
  }
    \begin{enumerate}
      \item \added{$\bar{\vec{z}}^t$ satisfies all CaVe-Primal constraints;}
      \item \added{$\sum_{s,c,v}U_s(X_v) {z}_{s,c,v}^* - \sum_{s,c,v}
        U_s(X_v)  \bar{z}_{s,c,v}^t \le \varepsilon$.}
    \end{enumerate}
  \label{thm:z}
\end{theorem}
\iftoggle{arxiv}{
\begin{IEEEproof}
\new
For 1), first it is straightforward to see that $\vec{z}^k$ for any $k$
satisfies the constraints~\eqref{eq:NUM8}, \eqref{eq:NUM19}, and \eqref{eq:NUM10},
since they are not relaxed when formulating \cave-Lagrangian and
\cave-User$_s$. Hence, it is easy to show that $\bar{\vec{z}}^t$ satisfies these three
constraints. As for the remaining constraint \eqref{eq:NUM9},
Theorem~\ref{thm:cave-dual} shows that for any $\varepsilon > 0$, there exists
an integer $K$ such that for every integer $t>K$ and for all $l$,
$|\lambda_l^{t}-\lambda_l^*|<\varepsilon/2$, where $\vec{\lambda}^*$ is the optimal point of CaVe-Dual.
Hence, for any integer $T>0$, $\lambda_l^{t+T+1}<\lambda_l^{t+1}+\varepsilon$.
Meanwhile, we know from Alg.~\ref{AlgCVS-Link} that
$\lambda_l^{k+1}\ge \lambda_l^{k} + h_k (\sum_{s,c,v}X_vH^l_{s,c}z_{s,c,v}^k-R_l)$.
Therefore,
$\lambda_l^{t+T+1}\ge \lambda_l^{t+1} + \sum_{k=t+1}^{t+T} h_k (\sum_{s,c,v}X_vH^l_{s,c}z_{s,c,v}^k-R_l)$. So we have
\[
  \frac{\sum_{k=t+1}^{t+T} h_k (\sum_{s,c,v}X_vH^l_{s,c}z_{s,c,v}^k-R_l)}
  {\sum_{k=t+1}^{t+T}h_k} < \frac{\varepsilon}{\sum_{k=t+1}^{t+T} h_k}.
\]
That is,
\[
  \sum_{s,c,v}X_vH^l_{s,c}\frac{\sum_{k=t+1}^{t+T} h_k z_{s,c,v}^k}
  {\sum_{k=t+1}^{t+T}h_k} < R_l + \frac{\varepsilon}{\sum_{k=t+1}^{t+T} h_k}.
\]
Let $T\goto\infty$, and we have
\[
  \sum_{s,c,v}X_vH^l_{s,c} \bar{z}_{s,c,v}^t \le R_l,
\]
for all $l$. Hence, $\bar{\vec{z}}^t$ satisfies the constraint \eqref{eq:NUM9},
and 1) is proved.
Since we also have $\lambda_l^{t+T+1}>\lambda_l^{t+1}-\varepsilon$,
we know $\sum_{s,c,v}X_vH^l_{s,c} \bar{z}_{s,c,v}^t \ge R_l$, and thus
$\sum_{s,c,v}X_vH^l_{s,c} \bar{z}_{s,c,v}^t = R_l$.

To prove
$\sum_{s,c,v}U_s(X_v) {z}_{s,c,v}^* - \sum_{s,c,v} U_s(X_v)  \bar{z}_{s,c,v}^t \le \varepsilon$, we note that based on Theorem~\ref{theorem:CV_strong_duality},
$\sum_{s,c,v}U_s(X_v) {z}_{s,c,v}^* = D(\vec{\lambda}^*)$.
Because of Theorem~\ref{thm:cave-dual}, for any $\varepsilon > 0$, there exists
an integer $K$ such that for every integer $t>K$ and for all $l$,
$|D(\vec{\lambda}^{t})-D(\vec{\lambda}^*)|<\varepsilon$.
By definition,
\begin{align*}
  D(\vec{\lambda}^t) =& \sum_{s,c,v}U_s(X_v)z_{s,c,v}^t - \sum_l \lambda_l^t
    \left( \sum_{s,c,v}X_v H^l_{s,c}z_{s,c,v}^t - R_l \right)\\
                     =& \sum_{s,c,v}U_s(X_v)z_{s,c,v}^t - \sum_l \lambda_l^*
    \left( \sum_{s,c,v}X_v H^l_{s,c}z_{s,c,v}^t - R_l \right)\\
  & + \sum_l (\lambda_l^* - \lambda_l^t) \left( \sum_{s,c,v}X_v H^l_{s,c}z_{s,c,v}^t - R_l \right)
\end{align*}
We know that
$D(\vec{\lambda}^t) \ge D(\vec{\lambda}^*) - \varepsilon$,
$|\lambda_l^t - \lambda_l^*|<\varepsilon/2$, and
$\sum_{s,c,v}X_v H^l_{s,c}z_{s,c,v}^t - R_l$ is bounded for all $l,t$ since
$0\le z_{s,c,v}^t\le 1$.
Let $B:=\frac{\mathbb{|L|}}{2}\max_{\vec{z}^t,l,t} |\sum_{s,c,v}X_v H^l_{s,c}
z_{s,c,v}^t - R_l| + 1$.
We then have
\begin{align*}
  D(\vec{\lambda}^*) \le& \sum_{s,c,v}U_s(X_v)z_{s,c,v}^t \\
  &- \sum_l \lambda_l^* \left( \sum_{s,c,v}X_v H^l_{s,c}z_{s,c,v}^t - R_l \right) + \varepsilon B.
\end{align*}
Taking weighted average of both sides from $k=t+1$ to $t+T$, and then
letting $T\goto\infty$, we have
\begin{align*}
  D(\vec{\lambda}^*)\le& \sum_{s,c,v}U_s(X_v) \bar{z}_{s,c,v}^t \\
  &- \sum_l \lambda_l^* \left( \sum_{s,c,v}X_v H^l_{s,c}  \bar{z}_{s,c,v}^t - R_l \right) + \varepsilon B.
\end{align*}
Note that $\sum_{s,c,v}X_vH^l_{s,c} \bar{z}_{s,c,v}^t = R_l$. Therefore,
\[
  \sum_{s,c,v}U_s(X_v) {z}_{s,c,v}^* =D(\vec{\lambda}^*) \le \sum_{s,c,v}
  U_s(X_v) \bar{z}_{s,c,v}^t + \varepsilon B,
\]
and this concludes the proof of 2).
\end{IEEEproof}
}{
\begin{IEEEproof}
  \added{See \cite{sasikumar2019cache}.}
\end{IEEEproof}
}

%% file: cp.tex
\section{The Content Placement Problem (CoP)}	\label{section:cp}

We now discuss the content placement (CoP) problem, which entails deciding $p_{c,v}$, the indicator function that network cache $c$ stores video version $v$, for all $c$ and $v$. As discussed in Section~\ref{section:model}, a major challenge to our optimization problem (\ref{eq:NUM1})--(\ref{eq:NUM6}) is that the vector $\vec{p}$ needs to be updated much less frequently than the vector $\vec{z}$. To address this challenge, we introduce pseudo-variables $\vec{z}':=[z'_{s,c,v}]$ and $\vec{p}':=[p'_{s,c,v}]$, which can be updated much more frequently than $\vec{p}$, to replace $\vec{z}$ and $\vec{p}$.\footnote{The pseudo-variables carry state information that needs to be shared between user applications and the network in the implementation.} We only update $\vec{p}$, the real content placement, after $\vec{p}'$ converges. Also, we relax (\ref{eq:NUM6}) by allowing $p'_{c,v}$ and $z'_{s,c,v}$ to be any real number between 0 and 1. We can now rewrite (\ref{eq:NUM1})--(\ref{eq:NUM6}) as:

\textbf{CoP-Primal}
\begin{maxi!}[2] %
{}{\sum_{s,c,v\in \mathbb{I}_s}U_s(X_v)z'_{s,c,v} 	\label{eq:cp-primal1}}{}{}
\addConstraint{\sum_vY_vp'_{c,v}}{\leq B_c,\label{eq:cp-primal2}}{\forall c\in\mathbb{C}}
\addConstraint{\sum_{c,v\in\mathbb{I}_s}z'_{s,c,v}}{=1,\label{eq:cp-primal3}}{\forall s\in\mathbb{S}}
\addConstraint{z'_{s,c,v}}{\leq p'_{c,v},\label{eq:cp-primal4}}{\forall s,c,v}
\addConstraint{\sum_{s,c,v}X_vH^l_{s,c}z'_{s,c,v}}{\leq R_l,\label{eq:cp-primal5}}{\forall l\in\mathbb{L}}
\addConstraint{0\leq p'_{c,v}\leq 1, 0\leq z'_{s,c,v}}{\leq 1,\quad\label{eq:cp-primal6}}{\forall s,c,v.}
\end{maxi!}

\subsection{Overview of the Solution}

Similar to our solution to the CaVe problem, we will consider a dual problem to the CoP-Primal problem. Let $\vec{\mu}':=[\mu'_{s,c,v}]$, and $\vec{\lambda}':=[\lambda'_l]$ be the vectors of Lagrange multipliers associated with each constraint in (\ref{eq:cp-primal4}) and (\ref{eq:cp-primal5}) respectively. The Lagrangian is then
\begin{align}
&L'(\vec{p}', \vec{z}', \vec{\lambda}', \vec{\mu}')\nonumber\\
&:= \sum_{s,c,v\in \mathbb{I}_s}U_s(X_v)z'_{s,c,v}-\sum_l\lambda'_l\big(\sum_{s,c,v}X_vH^l_{s,c}z'_{s,c,v}- R_l\big)\nonumber\\
&\phantom{:=}-\sum_{s,c,v}\mu'_{s,c,v}(z'_{s,c,v}-p'_{c,v}).
\end{align}

The dual objective, $D'(\vec{\lambda}', \vec{\mu}')$, is defined as the maximum value of $L'(\vec{p}', \vec{z}', \vec{\lambda}', \vec{\mu}')$ over $\vec{p}'$ and $\vec{z}'$ subject to constraints (\ref{eq:cp-primal2}), (\ref{eq:cp-primal3})  and (\ref{eq:cp-primal6}). We call the optimization problem CoP-Lagrangian:

\textbf{CoP-Lagrangian}
\begin{maxi!}[2]
{}{L'(\vec{p}', \vec{z}', \vec{\lambda}', \vec{\mu}')	\label{eq:CP-L1}}{}{}
\addConstraint{\sum_vY_vp'_{c,v}}{\leq B_c,\label{eq:CP-L2}}{\forall c\in\mathbb{C}}
\addConstraint{\sum_{c,v\in\mathbb{I}_s}z'_{s,c,v}}{=1,\label{eq:CP-L4}}{ \forall s\in\mathbb{S}}
\addConstraint{0\leq p'_{c,v}\leq 1,\ 0\leq z'_{s,c,v}\leq 1,}{\quad\label{eq:CP-L3}}{\forall s,c,v.}
\end{maxi!}

\begin{remark}
We note that an important difference between CoP-Lagrangian and CaVe-Lagrangian is that CoP-Lagrangian relaxes the constraint (\ref{eq:cp-primal4}) as well. Since the pseudo-variable $z'_{s,c,v}$ in CoP-Primal bears no physical meaning, this constraint can now be temporarily violated in practice.%
\end{remark}

The dual problem, which we call CoP-Dual, is to find the Lagrange multipliers that minimize $D'(\vec{\lambda}', \vec{\mu}')$:

\textbf{CoP-Dual}
\begin{mini!}
{}{D'(\vec{\lambda}', \vec{\mu}')\label{eq:CP-D1}}{}{}
\addConstraint{\lambda'_l }{\geq 0, \label{eq:CP-D2}}{\forall l \in \mathbb{L}}
\addConstraint{\mu'_{s,c,v}}{\geq 0,\quad}{\forall s\in\mathbb{S}, c\in\mathbb{C}, v\in\mathbb{V}.}
\end{mini!}

It is straightforward to show the following theorem:
\begin{theorem}
CoP-Primal and CoP-Dual have the same optimal value.
\end{theorem}
\iftoggle{arxiv}{
\begin{IEEEproof}
\added{We use the same justification as in the previous section.
  The objective
function of \cop-Primal is a linear function, and hence is concave. The set of
$\vec{z}'$ and $\vec{p}'$ that satisfies the three unrelaxed constraints,
namely, (\ref{eq:cp-primal2}),
(\ref{eq:cp-primal3}), and (\ref{eq:cp-primal6}), is nonempty and convex.

Furthermore, the
relaxed constraints (\ref{eq:cp-primal4}) and (\ref{eq:cp-primal5}) are
linear and thus convex. To get strict inequalities in
(\ref{eq:cp-primal4}) and (\ref{eq:cp-primal5}), we
i) choose $0< \varepsilon < \min\{ \frac{\min_l R_l}{|\mathbb{S}| |\mathbb{V}|}
  \frac{\min_s |\mathbb{I}_s|-1}{\max_v X_v}, \frac{\min_c B_c}{2 \sum_v
Y_v}\}$;
ii) set $p'_{c,v}$ to be $1$ if $c$ is the root node and $v$ is the null
version, and $2\varepsilon$ otherwise;
iii) set $z'_{s,c,v}$ to be $1-\varepsilon$ if $c$ is the root node and $v$ is
the null version, and $\frac{\varepsilon}{|\mathbb{C}| (|\mathbb{I}_s|-1)}$ otherwise.
It is straightforward to see that
\eqref{eq:cp-primal2},
\eqref{eq:cp-primal3}, and
\eqref{eq:cp-primal6} are satisfied, while
\eqref{eq:cp-primal4} and \eqref{eq:cp-primal5} are satisfied with strict
inequalities.

Hence, this theorem holds
following Theorem 6.2.4 (Strong Duality Theorem) in \cite{bazaraa2013nonlinear}.}
\end{IEEEproof}
}{
\begin{IEEEproof}
  \added{See \cite{sasikumar2019cache}.}
\end{IEEEproof}
}

We will solve CoP-Primal by solving CoP-Dual. %
We discuss our solutions to CoP-Lagrangian and CoP-Dual below.

\subsection{The Solution to CoP-Lagrangian}
We first rewrite $L'(\vec{p}', \vec{z}', \vec{\lambda}', \vec{\mu}')$ as:
\begin{align}
&L'(\vec{p}', \vec{z}', \vec{\lambda}', \vec{\mu}')\nonumber\\
&=\sum_s \sum_{c,v}z'_{s,c,v}\left(U_s(X_v)-X_v\sum_{l:H^l_{s,c}=1}\lambda'_l-\mu'_{s,c,v}\right) \nonumber\\
&\phantom{=}+\sum_{c}\sum_{v}p'_{c,v}\sum_s\mu'_{s,c,v}+\sum_l\lambda'_lR_l.
\end{align}

Let $\vec{z}'_s$ be the vector containing all $[z'_{s,c,v}]$ for a given $s$ and $\vec{p}'_c$ be the vector containing all $[p'_{c,v}]$ for a given $c$. Also, let $\bar{L}_s(\vec{z}'_s, \vec{\lambda}', \vec{\mu}'):=\sum_{c,v}z'_{s,c,v}[U_s(X_v)-X_v\sum_{l:H^l_{s,c}=1}\lambda'_l-\mu'_{s,c,v}]$, $\hat{L}_c(\vec{p}'_c,\vec{\mu}'):=\sum_{v}p'_{c,v}(\sum_s\mu'_{s,c,v})$, and $B(\vec{\lambda}'):=\sum_l\lambda_lR_l$. Then, we have 
\begin{align}
&L'(\vec{p}', \vec{z}', \vec{\lambda}', \vec{\mu}') \nonumber\\
&=\sum_s\bar{L}_s(\vec{z}'_s, \vec{\lambda}', \vec{\mu}')+\sum_c\hat{L}_c(\vec{p}'_c,\vec{\mu}')+B(\vec{\lambda}'),
\end{align}
which gives rise to a natural decomposition among all users and network caches. Specifically, consider the two subproblems, namely, CoP-User$_s$ and CoP-Cache$_c$, below. For fixed vectors $\vec{\lambda}'$ and $\vec{\mu}'$, CoP-Lagrangian can be solved by solving CoP-User$_s$ for each $s$ and CoP-Cache$_c$ for each $c$.

\textbf{CoP-User$_s$}
\begin{maxi!}
{}{\sum_{c,v}z'_{s,c,v}\big(U_s(X_v)-X_v\sum_{l:H^l_{s,c}=1}\lambda'_l-\mu'_{s,c,v}\big)}{}{}
\addConstraint{\sum_{c,v\in\mathbb{I}_s}z'_{s,c,v}}{=1}{}
\addConstraint{0\leq z'_{s,c,v}}{\leq 1,\quad}{\forall c\in\mathbb{C}, v\in\mathbb{V}.}
\end{maxi!}
\indent\textbf{CoP-Cache$_c$}
\begin{maxi!}
{}{\sum_{v}p'_{c,v}\sum_s\mu'_{s,c,v}}{}{}
\addConstraint{\sum_v Y_vp'_{c,v}}{\leq B_c}{}
\addConstraint{0\leq p'_{c,v}}{\leq 1,\quad}{\forall v\in\mathbb{V}.}
\end{maxi!}

CoP-User$_s$ can be solved by the following algorithm: First, find $(c^*,v^*)$ that has the maximum value of $U_s(X_v)-X_v\sum_{l:H^l_{s,c}=1}\lambda'_l-\mu'_{s,c,v}$ among all $(c,v)$ with $v\in \mathbb{I}_s$. Ties can be broken arbitrarily. Second, set $z'_{s,c^*,v^*}=1$, and $z'_{s,c,v}=0$ for all other $(c,v)$. Alg. \ref{AlgCP-User} shows the algorithm.

On the other hand, CoP-Cache$_c$ can be solved by the following greedy
algorithm: First, sort all video versions $v$ in decreasing order of
$\frac{\sum_s\mu'_{s,c,v}}{Y_v}$ so that $\frac{\sum_s\mu'_{s,c,1}}{Y_1}\geq
\frac{\sum_s\mu'_{s,c,2}}{Y_2}\geq\dots$. Second, starting from $v=1$, set
$p_{c,v}$ to be the largest possible value without violating any constraints.
Specifically, set $p'_{c,v}=\min\{1,(B_c-\sum_{v'<v}Y_{v'}p'_{c,v'})/Y_v\}$.
It is straightforward to verify that this greedy algorithm achieves the
optimal solution for CoP-Cache$_c$\added{, since it is a fractional knapsack
problem.}

\begin{remark}
Recall that $p_{c,v}$ is the indicator function that $c$ stores $v$, which needs to be an integer. The optimal solution to CoP-Cache$_c$ may not be integer. However, from the description of our greedy algorithm, it is obvious that, for each $c$, there is at most one $v$ with non-integer $p_{c,v}$. In practice, we make each network cache $c$ store only video versions with $p_{c,v}=1$. Since all but one version have integer $p_{c,v}$, this approach is close to optimal.
\end{remark}

\subsection{The Solution to CoP-Dual}
The CoP-Dual problem involves two Lagrange multipliers, $\vec{\lambda}'$ and $\vec{\mu}'$. They are updated as in Alg. \ref{AlgCP-Link} and \ref{AlgCP-Cache}. The following lemma and theorem, whose proofs are omitted due to space constraint, show that these algorithms solve CoP-Dual.

\begin{lemma}
  \label{lem:cop-subgradient}
Given $\vec{\lambda}'$ and $\vec{\mu}'$, let $\vec{z}'^*$ and $\vec{p}'^*$ be
the vectors that solve CoP-User$_s$ and CoP-Cache$_c$. Then the vector
\added{$\vec{g}':=[[R_l-\sum_{s,c,v}X_vH^l_{s,c}z'^*_{s,c,v}],
[p'^*_{c,v}-z'^*_{s,c,v}]]$} is a subgradient of $D'(\vec{\lambda}', \vec{\mu}')$.
\end{lemma}
\iftoggle{arxiv}{
\begin{IEEEproof}
  \new
    Since $\vec{z}'^*$ and $\vec{p}'^*$ solves CoP-User$_s$ and CoP-Cache$_c$
    respectively, they jointly solve CoP-Lagrangian for the given
    $\vec{\lambda}'$ and $\vec{\mu}'$. That is,
    $D'(\vec{\lambda}',\vec{\mu}') = L'(\vec{p}'^*,\vec{z}'^*,\vec{\lambda}',
    \vec{\mu}')$.
  Therefore, for any $\tilde{\vec{\lambda}}':=[\tilde{\lambda}'_l]$
  and $\tilde{\vec{\mu}}':=[\tilde{\mu}'_{s,c,v}]$,
  where $\tilde{\lambda}_l\ge 0$ and $\tilde{\mu}'_{s,c,v}\ge 0$,
  \begin{align*}
    &D'(\tilde{\vec{\lambda}}',\tilde{\vec{\mu}}') - D'(\vec{\lambda}', \vec{\mu}')\\
    &\quad \ge L'(\vec{p}'^*,\vec{z}'^*,\tilde{\vec{\lambda}}', \tilde{\vec{\mu}}') -
         L'(\vec{p}'^*,\vec{z}'^*,\vec{\lambda}',\vec{\mu}')\\
    &\quad = - \sum_l (\tilde{\lambda}'_l-\lambda'_l)
                \left(\sum_{s,c,v} z'^*_{s,c,v} H^l_{s,c} X_v - R_l\right)\\
    &\quad\mathrel{\phantom{=}} - \sum_{s,c,v} (\tilde{\mu}'_{s,c,v} - \mu'_{s,c,v})
                      (z'^*_{s,c,v} - p'^*_{c,v})\\
    &\quad = [\tilde{\vec{\lambda}}'-\vec{\lambda}', \tilde{\vec{\mu}}' - \vec{\mu}']^T \vec{g}'.
  \end{align*}
  Hence, $\vec{g}'$ is a subgradient of $D'(\vec{\lambda}',\vec{\mu}')$.
\end{IEEEproof}
}{
\begin{IEEEproof}
  \added{See \cite{sasikumar2019cache}.}
\end{IEEEproof}
}
\begin{theorem}
Let $\{h_t\}$ be a sequence of non-negative numbers with $\sum_{t=0}^\infty h_t=\infty$ and $\lim_{t\rightarrow\infty}h_t=0$, then Alg. \ref{AlgCP-Link} and \ref{AlgCP-Cache} together solve CoP-Dual.
\end{theorem}
\iftoggle{arxiv}{
\begin{IEEEproof}
  \added{%
    The proof is virtually the same as that of Theorem~\ref{thm:cave-dual}.
    Note that the objective function of \cop-Dual is convex in
    $\vec{\lambda}'$ and $\vec{\mu}'$, and the feasible region is a nonempty, convex, closed
    subset of $\mathbb{R}^{|\mathbb{L}|+|\mathbb{S}||\mathbb{C}||\mathbb{V}|}$.
    With Lemma~\ref{lem:cop-subgradient} and the step size sequence specified in the
    theorem, Alg.~\ref{AlgCP-Link}  and \ref{AlgCP-Cache} together solve CaVe-Dual
    following Theorem~8.9.2 in~\cite{bazaraa2013nonlinear}.
  }
\end{IEEEproof}
}{%
\begin{IEEEproof}
  \added{See \cite{sasikumar2019cache}.}
\end{IEEEproof}
}

\begin{algorithm}[h]
\caption{CoP-User$_s$ Algorithm}
\label{AlgCP-User}
\begin{algorithmic}[1]
  \STATE Obtain $\vec{\mu}'$ and $\vec{\lambda}'$
  \STATE $z'_{s,c,v}\leftarrow 0, \forall c,v$
  \STATE $(c^*,v^*) \leftarrow \arg\!\max_{c,v\in\mathbb{I}_s}U_s(X_v)-X_v\sum_{l:H^l_{s,c}=1}\lambda'_l-\mu'_{s,c,v}$
  \STATE $z'_{s,c^*,v^*} \leftarrow 1$
\end{algorithmic}
\end{algorithm}

\begin{algorithm}[h]
\caption{CoP-Link$_l$ Algorithm}
\label{AlgCP-Link}
\begin{algorithmic}[1]
\STATE $t\leftarrow 0$, $\lambda'_l \leftarrow 0$
\WHILE{true}  
\STATE Obtain $\vec{z}'$ from Alg. \ref{AlgCP-User}
\STATE $\lambda'_l \leftarrow \left[ \lambda'_l + h_t(\sum_{s,c,v}X_vH^l_{s,c}z'_{s,c,v} - R_l) \right]^+$ 
  \STATE $t\leftarrow t+1$
\ENDWHILE
\end{algorithmic}
\end{algorithm}

\begin{algorithm}[h]
\caption{CoP-Cache$_c$ Algorithm}
\label{AlgCP-Cache}
\begin{algorithmic}[1]
\STATE $t\leftarrow 0$, $\mu'_{s,c,v} \leftarrow 0$
\WHILE{true} 
\STATE Obtain $\vec{z}'$ from Alg. \ref{AlgCP-User}
\STATE $\mu'_{s,c,v} \leftarrow \left[\mu'_{s,c,v} + h_t(z'_{s,c,v}-p'_{c,v}) \right]^+\forall s, v$
\STATE Sort all versions so that $\frac{\sum_s\mu'_{s,c,1}}{Y_1}\geq \frac{\sum_s\mu'_{s,c,2}}{Y_2}\geq\dots$
\STATE $B'\leftarrow B_c$
\FOR{$v=1 \to |\mathbb{V}|$}
\STATE $p'_{c,v}\leftarrow \min\{1,\frac{B'}{Y_v}\}$
\STATE $B'\leftarrow B'-Y_vp'_{c,v}$
\ENDFOR
\STATE $t\leftarrow t+1$
\ENDWHILE
\end{algorithmic}
\end{algorithm}

\iftoggle{arxiv}{
\subsection{\added{The Solution to CoP-Primal}}

\added{We have the following theorem regarding the optimality of $\vec{z}'$
  obtained by running
Alg.~\ref{AlgCP-User}, Alg.~\ref{AlgCP-Link}, and Alg.~\ref{AlgCP-Cache} iteratively:}

\begin{theorem}
  \added{Let $\vec{z}'^*$ be the vector that solves CoP-Primal,
    $\vec{\lambda}'^k$ be the vector produced by Alg.~\ref{AlgCP-Link} after $k$ iterations,
    $\vec{\mu}'^k$ be the vector produced by Alg.~\ref{AlgCP-Cache} after $k$ iterations,
    and $\vec{z}'^k$ be the vector produced by Alg.~\ref{AlgCP-User} when
    $\vec{\lambda}'=\vec{\lambda}'^k$ and $\vec{\mu}'=\vec{\mu}'^k$.
    Let $\bar{\vec{z}}'^t$ be the weighted average of $\vec{z}^k$ after the
    first $t$ iterations, i.e.
    $\bar{\vec{z}}'^t := \lim_{T\goto\infty} \frac{\sum_{k=t+1}^{t+T} h_k
    \vec{z}'^k}{\sum_{k=t+1}^{t+T} h_k}$.
    Then, for any $\varepsilon > 0$, there exists an integer $K$ such that for
    every $t>K$,
  }
    \begin{enumerate}
      \item \added{$\bar{\vec{z}}'^t$ satisfies all CoP-Primal constraints;}
      \item \added{$\sum_{s,c,v}U_s(X_v) {z}'^*_{s,c,v} - \sum_{s,c,v}
        U_s(X_v) \bar{z}'^t_{s,c,v} \le \varepsilon$.}
    \end{enumerate}
  \label{thm:z-cp}
\end{theorem}
\begin{IEEEproof}
  \added{The proof is virtually the same as that of Theorem~\ref{thm:z}
  and thus omitted.}
\end{IEEEproof}
}{}

%% file: implementation.tex
\section{Implementation on Named Data Networking}	\label{section:implementation}

In this section, we discuss the implementation of our algorithms on Named Data Networking (NDN). We first introduce
the NDN architecture briefly, and then show how we implement our algorithms following the NDN philosophy.

\subsection{NDN Architecture}
NDN is a future Internet architecture where every piece of data is associated with a unique hierarchical name. When a user wants to obtain a piece of named data, the user device sends out an \emph{interest packet} with the name of the data. Note that usually the interest packet does not specify the destination location. NDN routers have built-in caches. When a router receives an interest packet, it first checks whether the named data is cached or not. If cached, it directly replies with the corresponding data packet.  Otherwise, it forwards the interest packet to the next hop according to the employed forwarding strategy. The content producer e.g. video service provider is responsible for generating data packets for a certain name space. The data packet follows the reverse route of the interest packet to the user.

\subsection{Placement of Data}\label{sec:data}
In our implementation, there are three types of data: packets of video contents, decision variables ($z'_{s,c,v}$ and $p_{c,v}$), and Lagrange multipliers ($\lambda_l$, $\lambda'_l$, and $\mu'_{s,c,v}$). We assign each of them a unique name. For example, a video version has a name prefix such as \verb|/r/file1/v1|, and $\mu'_{1,2,3}$ has \verb|/mu2/1_3|. Each prefix is appended a sequence number to uniquely identify video packets and variables in different iterations. Naturally, video contents are placed at network caches according to the video versions.\footnote{Videos are cached in full rather than at the packet level.} Decision variables $z'_{s,c,v}$ are stored and updated at the corresponding user $s$. Decision variables $p_{c,v}$ and Lagrange multipliers $\mu'_{s,c,v}$ are stored and updated at the corresponding network cache $c$. Finally, Lagrange multipliers $\lambda_l$ and $\lambda'_l$ of link $l$ from node $A$ to $B$ are stored and updated at node $A$ that is closer to the cache.

\subsection{Implementation of User Algorithms}\label{sec:impl-user}
From Alg.~\ref{AlgCVS-User} and \ref{AlgCP-User}, we can see that each user $s$ needs to know the values of $p_{c,v}$, $\lambda_l$, $\lambda'_l$, and $\mu'_{s,c,v}$. Each user periodically sends out interest packets for the named data of these variables. Since the names of these data indicate the entities that store them, routers can easily route the interest packets to the correct destinations. Further, as data packets traverse in the reverse route of their corresponding interest packets, each router can cache all latest values of $p_{c,v}$ and $\lambda_l$ that pass through it.

With the information of $p_{c,v}$ and $\lambda_l$, each user $s$ can find the best video version $v^*$ and cache $c^*$ via Alg.~\ref{AlgCVS-User}. User $s$ then sends out interest packets for video version $v^*$ at a rate indicated by $X_{v^*}$. Note that these interest packets only contain information about the video version $v^*$, and not the destination $c^*$. Nevertheless, the following forwarding strategy ensures the interest packet will be eventually forwarded to $c^*$ assuming no link failure or topology change: When a router receives an interest packet for video version $v^*$, it finds the network cache $c^\dag$ that has the smallest \emph{cost}, where the cost is defined as $\sum_l\lambda_l$ over all link $l$ on the path to the network cache $c$, among those that store $v^*$, i.e., $p_{c,v^*}=1$. It then forwards the interest packet to the next router on the path toward $c^\dag$. Note that routers store all values of $p_{c,v}$ and $\lambda_l$ that pass through it and thus do not need additional message passing.

With the information of $p_{c,v}$, $\lambda'_l$ and $\mu'_l$, each user $s$ can decide the video version $v^*$ and network cache $c^*$ such that $z'_{s,c^*,v^*}=1$ via Alg.~\ref{AlgCP-User}. Each user $s$ then sends out a pseudo-interest packet with the name of $z'_{s,c^*,v^*}$. We call it a pseudo-interest packet since it is used to inform the caches the changes of $z_{s,c,v}$. The replied data packet from cache $c^*$ carries no meaning payload and is ignored.

\subsection{Implementations for Routers and Caches}
We now discuss the implementations of Alg.~\ref{AlgCVS-Link}, \ref{AlgCP-Link}, and \ref{AlgCP-Cache}.
In Alg.~\ref{AlgCVS-Link}, each router needs to know $\sum_{s,c,v}X_vH^l_{c,v}z_{s,c,v}$ to update $\lambda_l$ for its links. We note that $\sum_{s,c,v}X_vH^l_{c,v}z_{s,c,v}$ can be estimated by the product of the rate of interest packets going through the opposite link to $l$ and video data packet size. As the router knows the rate of interest packets going through $l$, it can update $\lambda_l$ directly without requesting additional information. Likewise, Alg. \ref{AlgCP-Link}, and \ref{AlgCP-Cache} can be carried out if one knows $z'_{s,c,v}$. This is achieved by user $s$ sending out a pseudo-interest packet as explained in Section~\ref{sec:impl-user}. Besides, $R_l$, the maximum supportable data rate of link $l$, is obtained from stress tests.

%% file: simulation.tex
\section{Evaluations}
\label{sec:sim}

We present our simulation evaluation results in this section. All simulations are conducted on ndnSIM \cite{mastorakis2017ndnsim}, an ns-3 based NDN simulator.

We consider the wireless edge network in \figurename~\ref{fig:topo} for evaluation. Same as in~\cite{ramadan2017big}, the topology of network caches follows the three-tier hierarchy of the YouTube video delivery system. There are $15$ network routers with caches in total, including the root node and $8$ edge caches. Each edge cache serves $20$ users who have different types of devices and are interested in different videos.

We consider a catalog of $200$ different videos, each with $5$ different versions. The popularity of these videos follows the Zipf distribution with the shape parameter equal to $1$. The $5$ versions correspond to video resolutions of 360p, 720p, 1080p, 1440p (2K), and 2160p (4K) respectively.\footnote{The aspect ratio is assumed to be $16:9$ as in YouTube. For example, a 720p video has a resolution of $1280\times 720$.} The data rate of streaming each video version is set based on measurement results for YouTube videos with H.264 codec~\cite{mcfly2018test}. The access link capacities between users and edge caches are \SI{25}{Mbps} each so that one can stream a 4K video. The capacities of links between caches and the root node are \SI{100}{Mbps} each so that the number of concurrent 4K streams is low. We assume each video is one-hour long, and the file sizes of video versions are calculated accordingly. The root node holds all video versions. Each edge (or primary), secondary, and tertiary cache is assumed to be able to hold all versions of one, two, and four videos respectively.

As for user utilities, we categorizes user devices into three types: smartphones, laptops or tablets, and TVs. The utility function of each user has the form $U(X_v) = \alpha \ln \min(X_v, \bar{X})$, where $\alpha$ is a scaling factor capturing the effect of the screen size, $X_v$ is the data rate of video version $v$ in \si{Mbps}, and $\bar{X}$ is a cutoff rate reflecting the limit of the device resolution.
For the three types, we set a scaling factor of $20, 40, 60$ and a cutoff rate corresponding to a 1080p, 2K, 4K video respectively.
Besides, \replaced{we set $U(0)=-100$}{when the user's video streaming stalls,
the user gets a null utility of $-100$}, which is much smaller than all regular utilities.

To evaluate the performance of our algorithms,
we implement and compare the following four policies:
\begin{itemize}
\item \textbf{Optimal}: This policy tries to find the optimal solution to the CaVe-CoP problem by solving the integer program numerically via the GLPK toolbox.
Note that it is a centralized policy and involves solving a high-dimensional problem.
\item \textbf{CaVe-CoP}: This refers to our algorithms Alg.~\ref{AlgCVS-User}--\ref{AlgCP-Cache}.
\item \textbf{CaVe-CAV}: In this policy, each user employs our algorithms for CaVe. For content placement, if a network cache decides to store a video version, it needs to cache all versions (CAV) of the same video. We note that this content placement strategy is consistent with design practices in commercial CDNs. As a result, each network cache simply stores the most popular videos, subject to its storage constraint.
\item \textbf{Greedy-CoP}: In this policy, each user chooses the version that matches its cutoff rate. Network caches employ our algorithms for CoP.
\end{itemize}

For each simulation, we use the video contents that each user actually
receives to calculate \deleted{both} the total utility \added{of all users} and the
\added{average} \% stall time, i.e. the percentage of time that video
streaming stalls\footnote{\added{Video streaming stalls when all
received video contents are consumed.}}\added{, of all users}.
The metrics are calculated at each \cave iteration, i.e. every \SI{0.1}{s}. We run \cop iterations every \SI{0.2}{s},
and apply content placement results at \SI{20}{s}.

\begin{figure}[!t]
  \centering
  \includegraphics[width=0.35\textwidth]{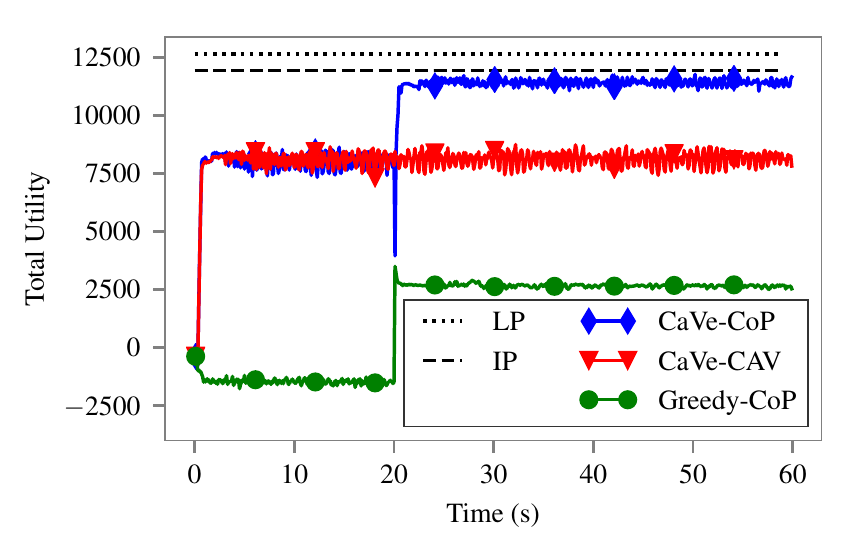}
   \caption{Comparison of total utility.}
   \label{fig:util}
\end{figure}

\begin{figure}[!t]
  \centering
  \includegraphics[width=0.35\textwidth]{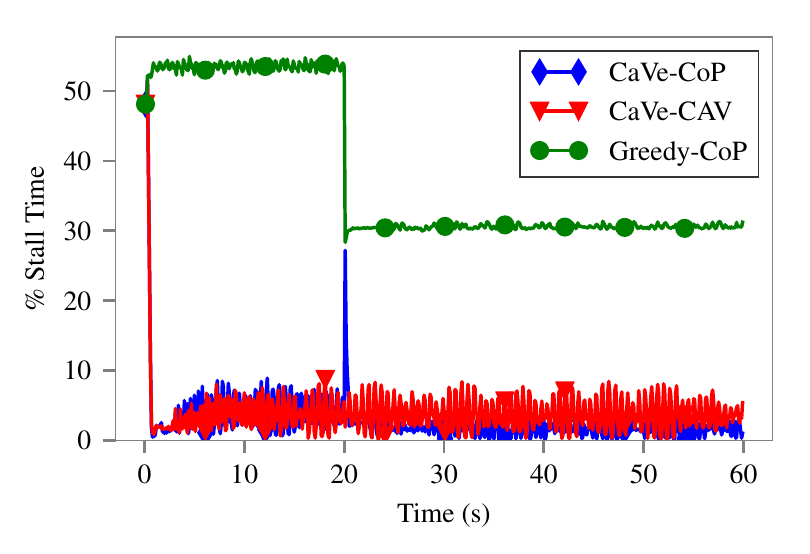}
   \caption{Comparison of \% stall time.}
   \label{fig:stall}
\end{figure}

Fig.~\ref{fig:util} and Fig.~\ref{fig:stall}
present our simulation results.
For our simulated scenario, Optimal cannot find the exact integer solution for utility.
Instead, it reports a upper bound from linear programming (LP) relaxation, and a lower bound by integer programming (IP) heuristics. Note that Optimal reports ideal utility instead of perceived utility.
We can observe that our Cave-CoP policy achieves
near-optimal utility, significantly outperforming the two baseline policies, even though they involve subsets of our algorithms.
Besides, our policy approaches zero stall time.
\added{Note that the jumps near \SI{20}{s} in the figures are due to applying content placement
results.}

%% file: related.tex
\section{Related Work}
\label{sec:related}

There has been rich literature on adaptive video streaming. An early work identified a cross layer framework for adaptive video streaming in IP networks~\cite{ahmed2005adaptive}.
More recently,
experiment-based investigations have been conducted on \replaced{YouTube}{the
YouTube video delivery system}~\cite{adhikari2012vivisecting}\replaced{ and
Netflix~\cite{adhikari2012unreeling}}{ as well as
CDNs run by Akamai~\cite{decicco2010experimental} and
Netflix~\cite{adhikari2012unreeling}}.
\added{Liu~\etal~\cite{liu2012case} made a case for a coordinated control
plane across CDNs for video streaming to provide high quality of experience.}
\deleted{The performance has also been investigated on
NDN~\cite{rainer2016investigating}.}
Wireless edge networks are promising to enhance the benefits of CDNs for
adaptive video streaming~\cite{tran2017collaborative}\deleted{, where the user
scheduling problem has been studied~\cite{bethanabhotla2014adaptive}}.
Content caching is crucial for practical adaptive video streaming in wireless edge networks. Considering distributed caches, Ramadan~\etal~\cite{ramadan2017big} proposed the abstraction of ``BIG'' cache to effectively utilize the resources.
\added{Applegate~\etal~\cite{applegate2016optimal} studied optimal content
placement of videos with a focus on scalability.}
There have been many studies on joint optimization of content caching and packet routing.
Yeh~\etal~\cite{yeh2014vip} proposed a framework for joint forwarding and caching in NDN. %
Wang~\etal~\cite{wang2018distributed} employed stochastic network utility maximization and developed a distributed forwarding and caching algorithm.
Ioannidis and Yeh~\cite{ioannidis2017jointly} studied the routing cost minimization problem of joint routing and caching, where the cost is incurred per link. These studies are not directly applicable to multi-version video streaming since different versions of the same video can be stored in different caches.%

\iftoggle{arxiv}{
Our work formulates the joint cache-version selection and content placement problem as a network utility maximization (NUM) problem, and uses the well-known primal dual approach and dual decomposition~\cite{palomar2006tutorial}. However, there are notable differences between our work and traditional NUM research. Existing studies have explored various scenarios including time varying channel with delay constraints~\cite{hou2010utility}, delay sensitive fairness~\cite{eryilmaz2017discounted},
multiple flow classes~\cite{gupta2016centralized},
multiple protocols~\cite{ramaswamy2014which} and so on, while assuming a static source-destination pair per user (flow). In contrast, in our work, a user could obtain its desired content from in-network caches as well as the content producer. %
}{}

%% file: conclusion.tex
\section{Conclusion}
\label{section:conclusion}

In this paper, we have studied the CaVe-CoP problem, i.e. the joint optimization of cache-version selection and content placement, for adaptive video streaming in wireless edge networks. Realizing that there is a practical timescale separation between CaVe and CoP, we have proposed a set of algorithms that provably optimize CaVe and CoP respectively.  Further, we show that our algorithms can be practically implemented on NDN in a distributed fashion.  Simulation evaluations on ndnSIM demonstrate that our policy significantly outperforms baseline policies with conventional heuristics.